\documentclass[12pt,a4paper,leqno]{article}
\usepackage{graphicx}
\title{Generalization of the model of conflict between two armed groups}
\author{Nikolay K. Vitanov$^{*}$ and S. Panchev$^{**,***}$}
\date{}
\begin{document}
\maketitle
\begin{abstract}
The conflicts between armed groups often go on for years. The classical
model of such conflicts accounts for the number of participants and for
the technology level of the equipment of the groups. Below we 
extend this model in order to account for events that are present for
limited time. As examples we discuss three kinds of such events: 
inclusion of reserves, presence of epidemics and use of non-conventional 
weapons. We show that if such events are not handled properly 
by the leaders of the groups the corresponding group can lose the conflict. 
\end{abstract}
\section{Introduction}
Population dynamics deals with coexistent animal or human populations [1-4].
The most antagonistic relations among animal populations
are the predator-prey ones. Human populations are different as
they can compete economically or politically [5-8]
or they can support fighting among armed groups [9-10].
Several decades ago Richardson and Lanchester applied the idea 
for mathematical modeling of arms races and military combats [11-12]
In our unstable world today such kind of modeling becomes highly actual [13-14]
\par
Usually the number of members and the quality of their equipment are the most important
characteristics of the armed groups. A general model of conflict between 
two such groups (called the "Red group" and the "Blue group") is  
\begin{equation}\label{model1}
\frac{dB}{dt}=F(B, R;b,r), \hskip.5cm
\frac{dR}{dt}=G(B, R;b;r)
\end{equation}
where $R(t)$ and  $B(t)$ are the  numbers of armed members of the two groups; 
$b$  and $r$ account for the technology level of the equipment 
of the two groups; and $F$ and $G$ are linear or nonlinear
functions, depending on the character of the conflict. 
Epstein \cite{e4} discussed the following particular case of 
the model (\ref{model1})
\begin{equation}\label{epst1}
\frac{dR}{dt}= - bB^{c_{1}} R^{c_{2}}, \hskip.5cm
\frac{dB}{dt}= - rR^{c_{3}} B^{c_{4}}
\end{equation}
where $c_{1,2,3,4}$ are  real nonnegative coefficients. 
\par
Below we generalize the model (\ref{model1}) by addition of terms
which describe the influence of different events acting for limited time.
In section 3 we discuss three simple cases of such events: inclusion of
reserves, epidemics and use of non-conventional weapons. Several
concluding remarks are summarized in section 4.
\section{Generalization of the model (\ref{model1})}
Let us introduce the function
\begin{equation}\label{switch}
V(t,t_{1},t_{2},\mu, \nu) = \Theta (t_{1}) \{ 1- \exp[\nu (t_{1}-t)]
\} \exp[\Theta(t_{2}) \nu (t_{2}-t)]
\end{equation}
where $t_{2}>t_{1}$ and $\Theta$ is the Heaviside theta function.
$V$ describes the following process: at $t_{1}$ $V$ grows from
$0$ to almost $1$ and this growth is controlled by the parameter $\mu$.
At $t=t_{2}$ $V$ begins an exponential decrease from $1$ to $0$. This decrease is 
controlled by the parameter $\nu$ (see Fig. 1).   
\begin{figure}[t]
\begin{center}
\includegraphics[angle=0,width=6.5cm]{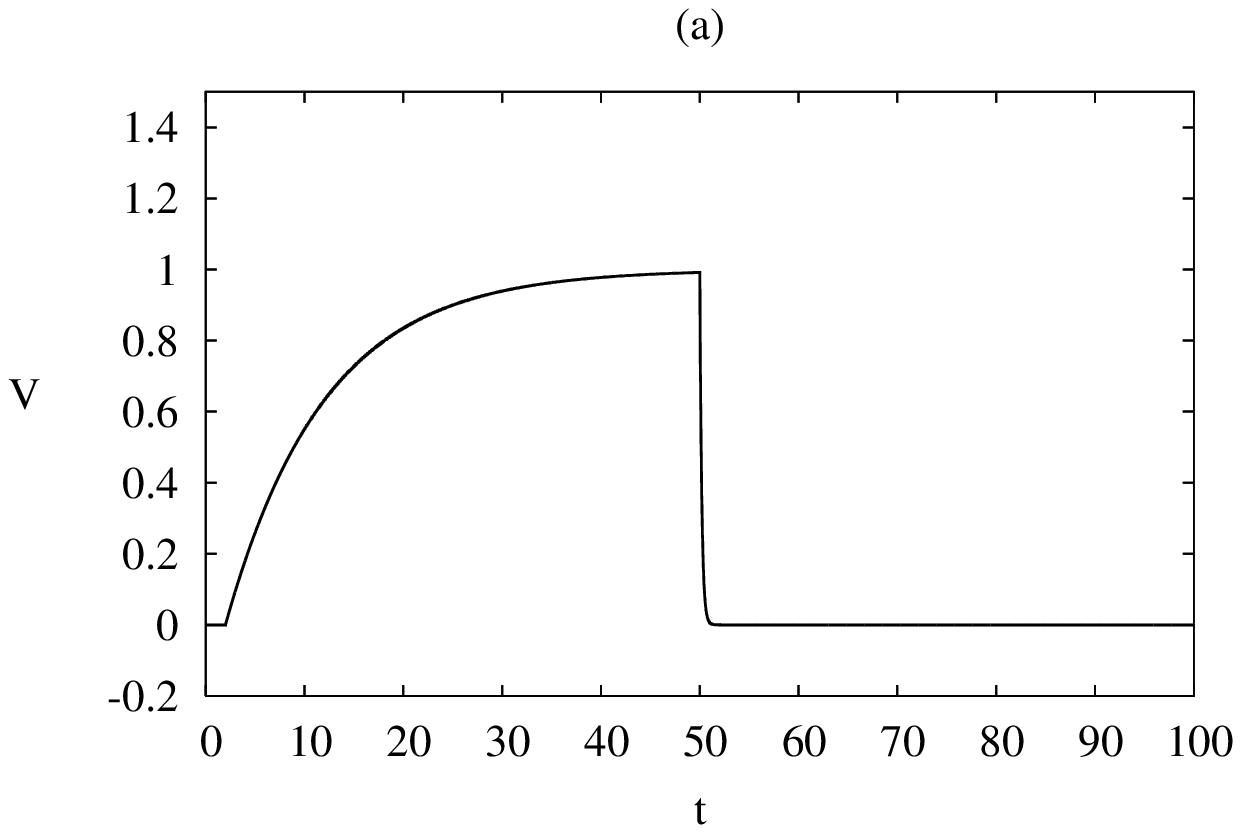}
\includegraphics[angle=0,width=6.5cm]{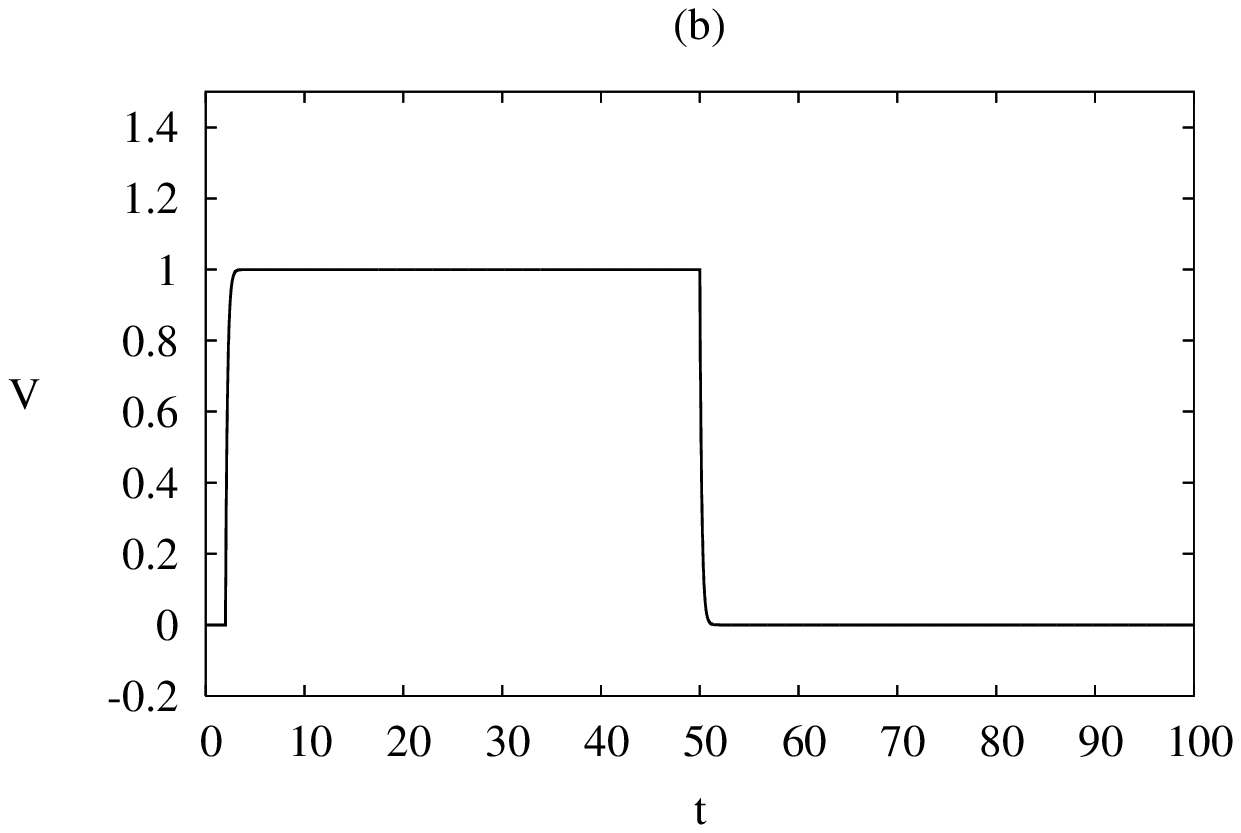}
\includegraphics[angle=0,width=6.5cm]{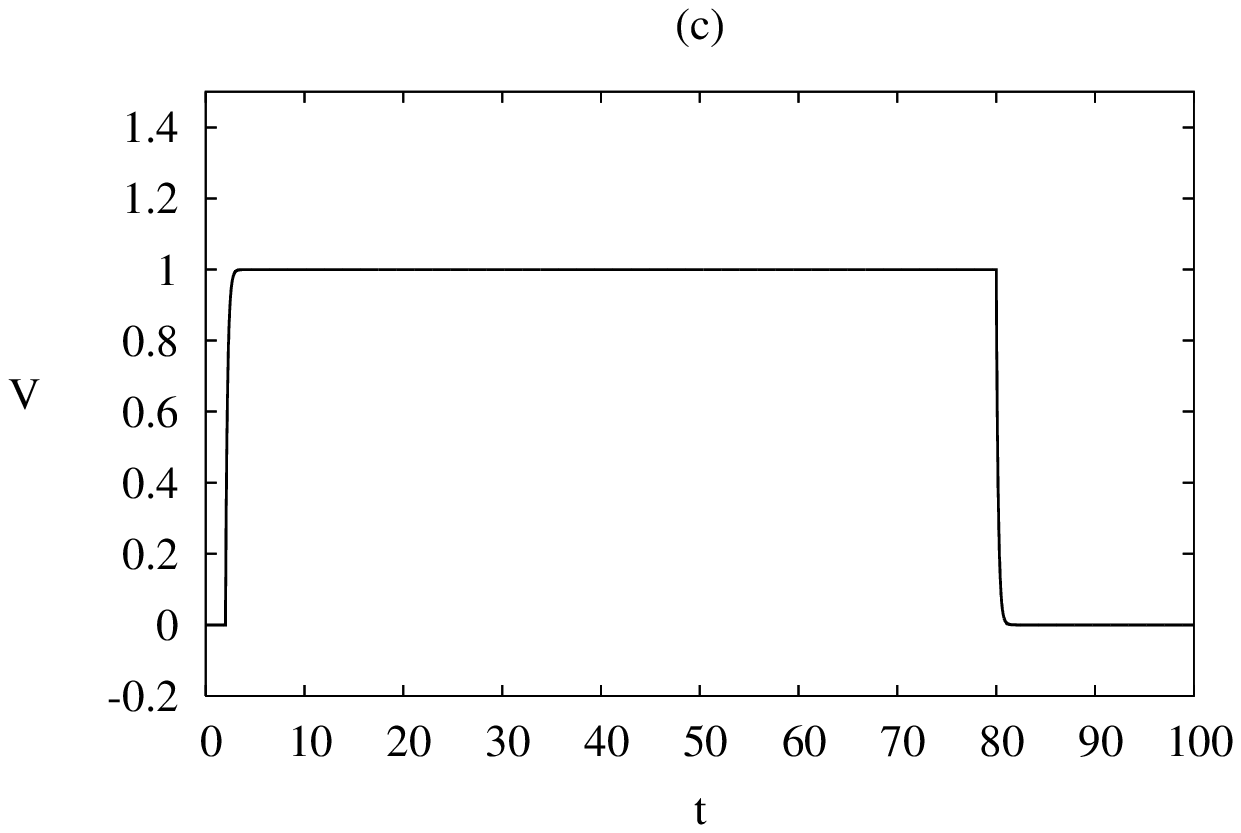}
\end{center}
\caption{The function $V(t,T_{1},t_{2},\mu,\nu)$.
Fig. 1a: long switch -on  fast switch-off ($\mu=0.1$, $\nu=5.0$). Fig. 1b:
fast switch-on, fast switch-off ($\mu=\nu=5.0$). $t_{1}=2.5$,$t_{2}=50$ Fig. 1c: fast switch-on,
fast switch-off, $t_{1}=2.5$, $t_{2}=80$.}
\end{figure}
\par
By means of the function $V$ we can include different effects that
act for limited  time. Three examples are
\begin{enumerate}
\item
Inclusion of reserves: $N_{1}$ times from  $t_{1}^{(i)}$ till
 $t_{2}^{(i)}$ for the Blue group (amplitude $B_{i}$) and 
$N_{4}$ times from  $t_{1}^{(p)}$ till
 $t_{2}^{(p)}$ for the Red group (amplitude $R_{p}$)
$$ 
\sum_{i=1}^{N_{1}} B_{i} V(t,t_{1}^{(i)},t_{2}^{(i)},\mu_{i}, \nu_{i}),
\hskip.5cm
\sum_{p=1}^{N_{4}} R_{p} V(t,t_{1}^{(p)},t_{2}^{(p)},\mu_{p}, \nu_{p})
$$
\item
Epidemics: $N_{2}$ times from $t_{1}^{(j)}$ till
 $t_{2}^{(j)}$ for the Blue group (amplitude $C_{j}$) and 
from $t_{1}^{(q)}$ till
 $t_{2}^{(q)}$ for the Red group (amplitude $E_{q}$)
$$ 
\sum_{j=1}^{N_{2}} C_{j} V(t,t_{1}^{(j)},t_{2}^{(j)},\mu_{j}, \nu_{j}),
\hskip.5cm
\sum_{q=1}^{N_{5}} E_{q} V(t,t_{1}^{(q)},t_{2}^{(q)},\mu_{q}, \nu_{q})
$$
\item
Using of non-conventional weapons: $N_{3}$ times from  $t_{1}^{(k)}$ till
 $t_{2}^{(k)}$ against the Blue group (amplitude $D_{k}$) and
$N_{6}$ times from  $t_{1}^{(v)}$ till
 $t_{2}^{(v)}$ against the Red group (amplitude $H_{v}$).
$$ 
\sum_{k=1}^{N_{3}} D_{j} V(t,t_{1}^{(k)},t_{2}^{(k)},\mu_{k}, \nu_{k}),
\hskip.5cm
\sum_{v=1}^{N_{6}} H_{v} V(t,t_{1}^{(v)},t_{2}^{(v)},\mu_{v}, \nu_{v})
$$
\end{enumerate}
$D_{k}$ and
 $H_{v}$ can depend on parameters characterizing the kind of the non-conventional
weapon.
\par
The general system of model equations becomes
\begin{eqnarray}\label{gensys1}
\frac{dB}{dt} = F(B,R;b,r) + \sum_{i=1}^{N_{1}} B_{i} V(t,t_{1}^{(i)},t_{2}^{(i)},\mu_{i}, \nu_{i})
-\nonumber \\ -\sum_{j=1}^{N_{2}} C_{j} V(t,t_{1}^{(j)},t_{2}^{(j)},\mu_{j}, \nu_{j})
-\sum_{k=1}^{N_{3}} D_{j} V(t,t_{1}^{(k)},t_{2}^{(k)},\mu_{k}, \nu_{k})
\end{eqnarray}
\begin{eqnarray}\label{gensys2}
\frac{dR}{dt} = G(B,R;b,r) + \sum_{p=1}^{N_{4}} R_{i} V(t,t_{1}^{(p)},t_{2}^{(p)},\mu_{p}, \nu_{p})
-\nonumber \\ -\sum_{q=1}^{N_{5}} E_{j} V(t,t_{1}^{(q)},t_{2}^{(q)},\mu_{q}, \nu_{q})
-\sum_{v=1}^{N_{6}} H_{j} V(t,t_{1}^{(v)},t_{2}^{(v)},\mu_{v}, \nu_{v}) 
\end{eqnarray} 
\section{Epidemics, reserves, non-conventional weapons}
\subsection{Epidemics in a position conflict}
Let us assume that an epidemics situation exists for the Blue group
from time $t_{1}$ to the time $t_{2}$. The system of model equations
become
\begin{equation}\label{epid1}
\frac{dB}{dt}= -rR - C V(t,t_{1},t_{2},\mu,\nu),
\hskip.5cm
\frac{dR}{dt}= -bR
\end{equation}
where we assume that the amplitude $C$ connected to the epidemics is
a constant. Figure 2 illustrates the danger of epidemics.
We have a Blue group that at $t=0$ is two times larger than the
Red group and we have an epidemics that starts at $t=20$ and begins
to decrease its death toll at $t=80$. At moderate ampltude of
epidemics - Fig. 2a - the Blue group still wins but if the
amplitude of epidemics is large and there are no effective
countermeasures the two times larger Blue group loses the conflict
as it can be seen in Fig. 2b.
\begin{figure}[h]
\begin{center}
\includegraphics[angle=0,width=6cm]{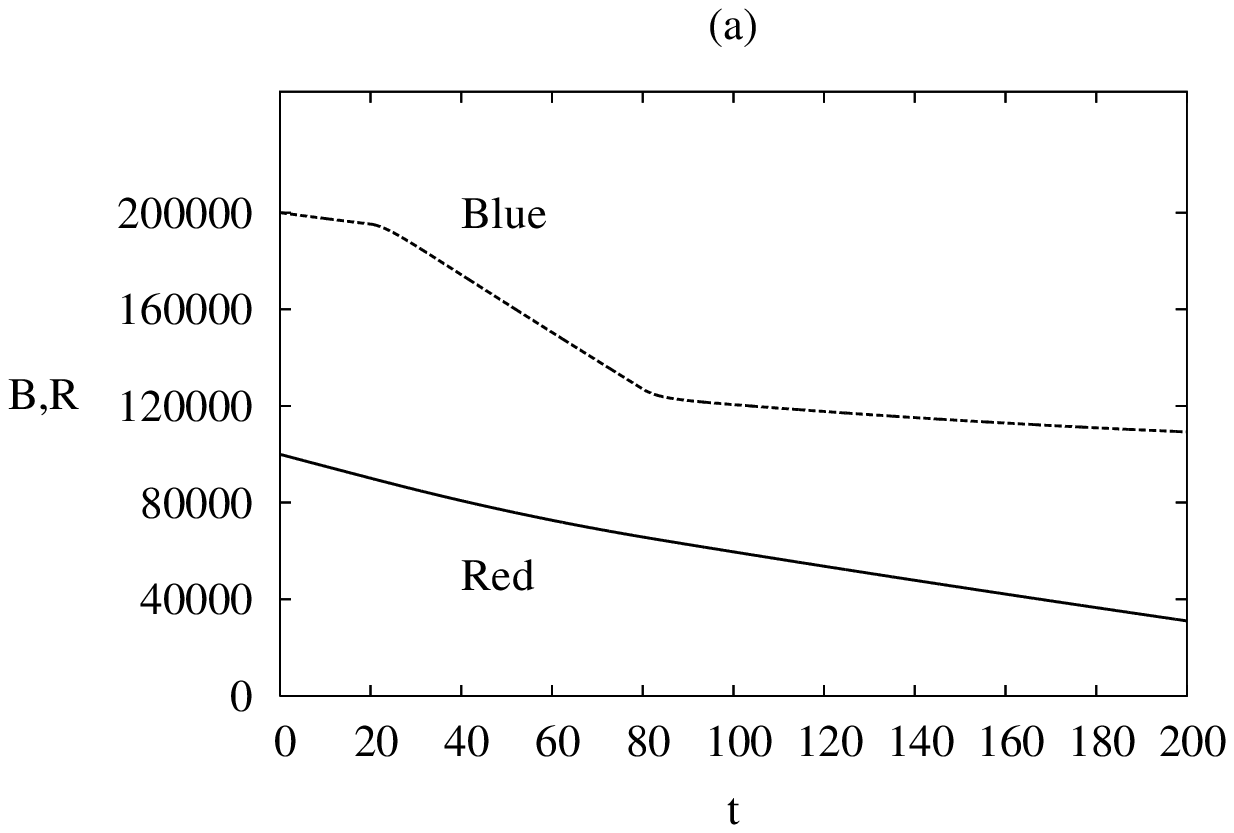}
\includegraphics[angle=0,width=6cm]{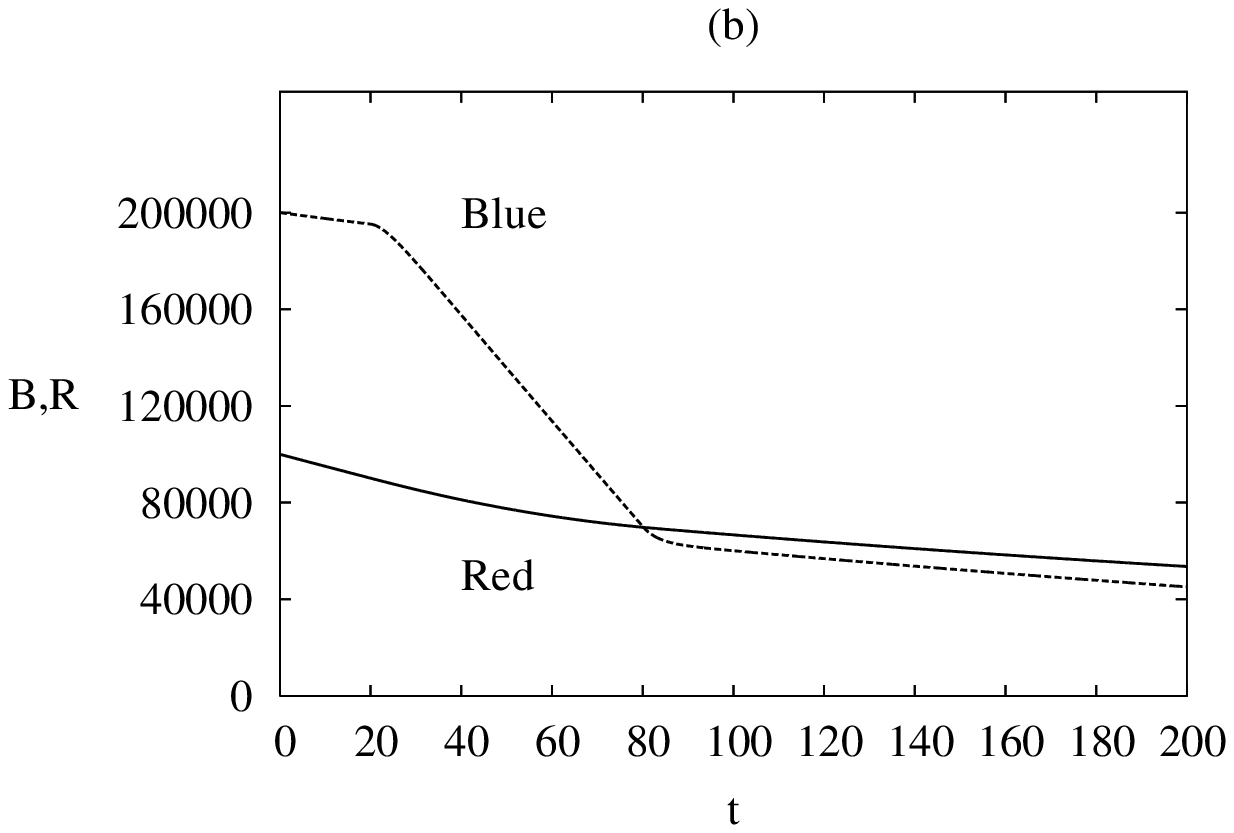}
\end{center}
\caption{Influence on an epidemics on the result of
a position conflict. At $t=0$ the Blue group is two times numerous as the
Red group: $B_{0}=200000$, $R_{0}=100000$. The epidemics starts
at $t=20$ and it begins do decay rapidly at $t=80$. $\mu = \nu =3$
and the firepower effectiveness of the two groups is the same
$b=r=0.0025$. Figure (a) : amplitude of disease $C=1000$ - Blue group still
wins. Figure (b): amplitude of disease $C=2000$ - the Blue group losses
the conflict.}
\end{figure} 
\subsection{Using reserves to counter an attack}
Now let a five time larger Red group attacks the Blue group
and the Blue group has some reserve members that can be used
in the conflict. We shall consider here the case of
including all reserves at once. The system of equations
becomes
\begin{equation}\label{reserves1}
\frac{dR}{dt} = -b R B, \hskip.5cm
\frac{dB}{dt} = -r R B + B_{1}V(t,t_{1},t_{2},\mu,\nu)
\end{equation}
The result from the attack depends on the quantity and the speed
of introduction of the reserves of the Blue group. Several
outcomes are presented in Fig. 3. If the reserves are
enough but are not introduced fast enough the attack of the Red group
is successful. If the reserves are introduced fast enough the
Blue group stops the attack and wins. These observations confirm the
existence of thresholds in  the nonlinear systems. 
If the change of the system is below the threshold
the system eliminates the perturbation (as in Fig. 3a). If the perturbation
is large enough however (over the threshold) the system state can
be changed (as in Fig. 3c).
\begin{figure}[h]
\begin{center}
\includegraphics[angle=0,width=6cm]{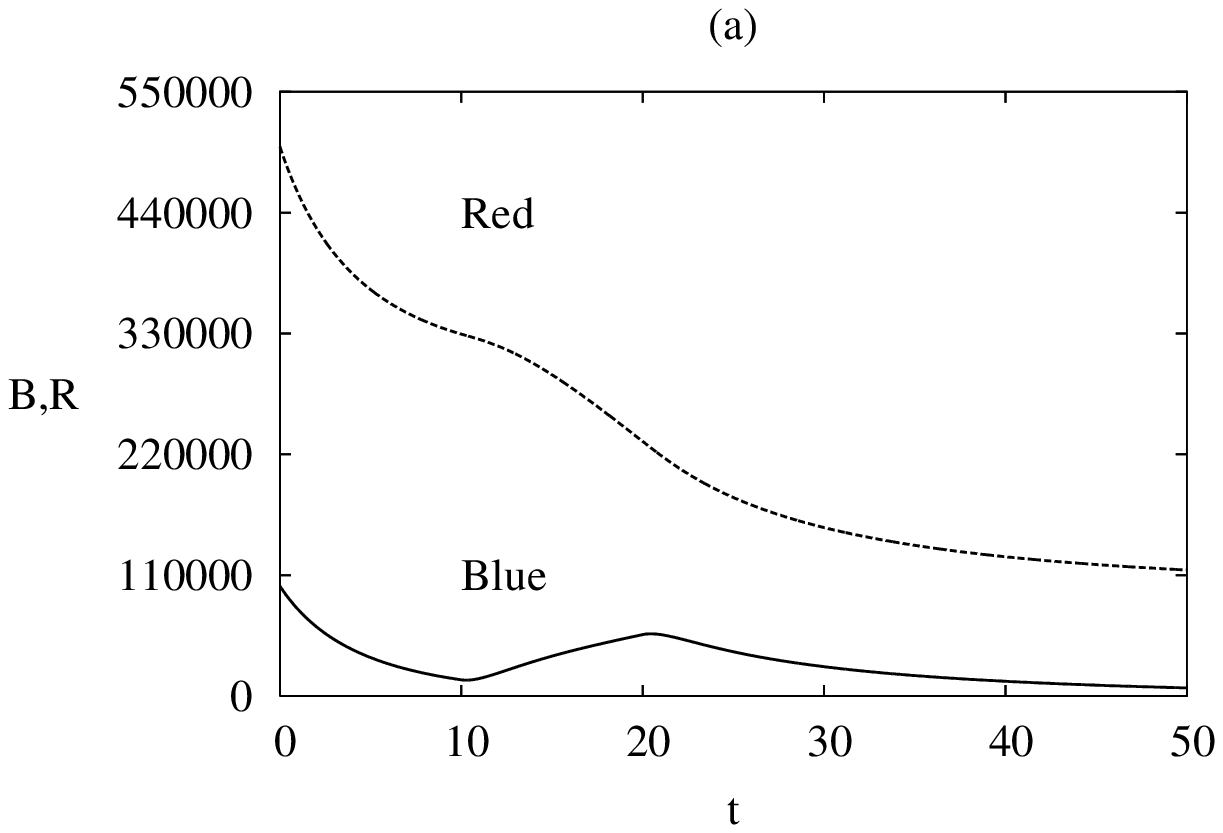}
\includegraphics[angle=0,width=6cm]{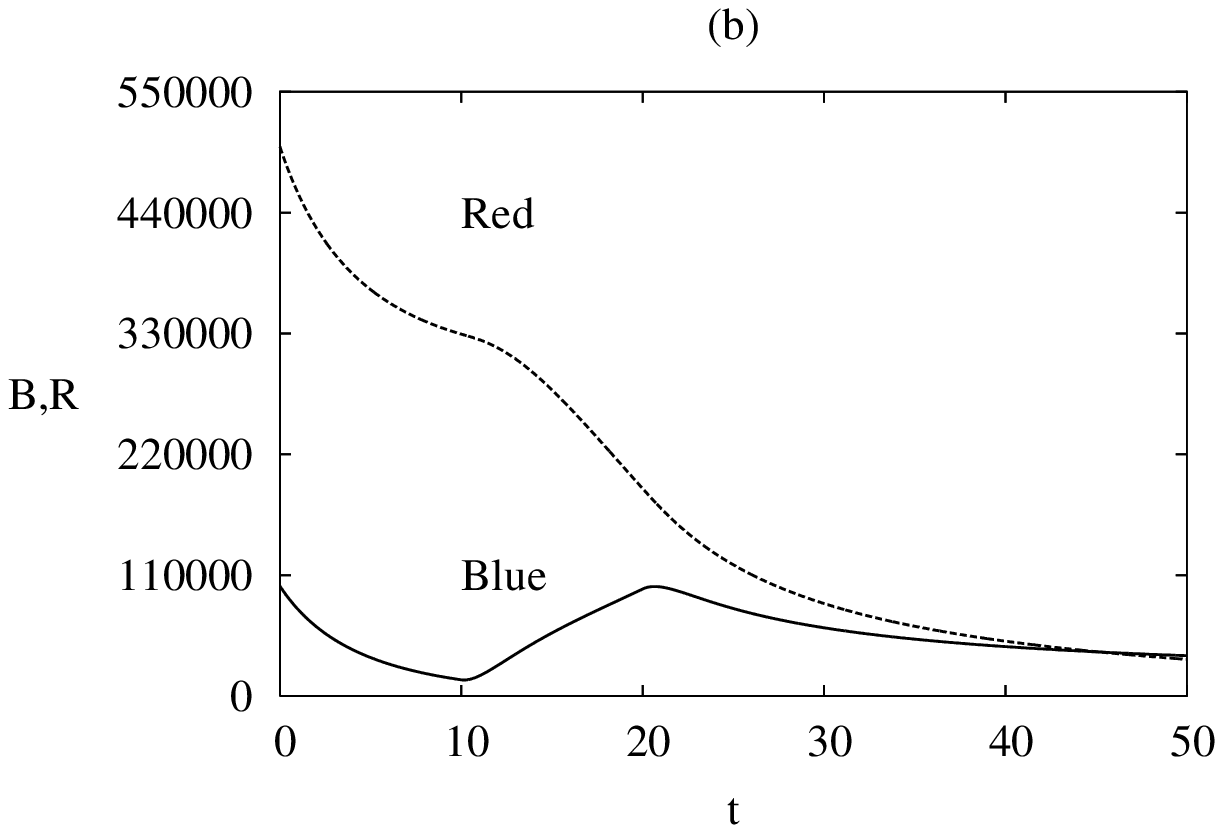}
\includegraphics[angle=0,width=6cm]{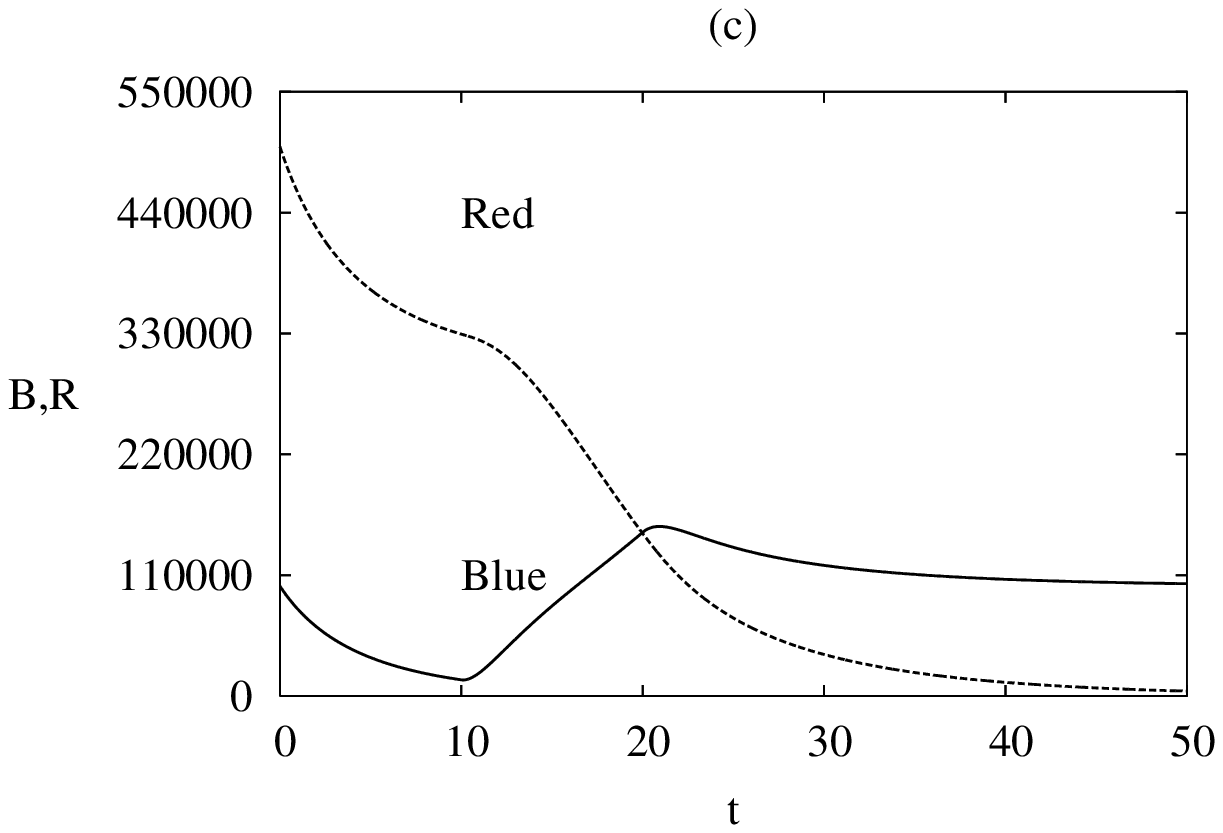}
\end{center}
\caption{Influence of reserves on an attack. Red group of 500 000 man attacks Blue 
group of 100 000 man. $b=10^{-6}$, $r=5 \cdot 10^{-7}$. Blue group
introduces reserves from $t=10$ till $t=20$. $\mu = \nu =1$. Figure
(a): amplitude $B_{1}=10^{4}$. The Red army group. Figure (b):
amplitude $B_{1}=1.7 \cdot 10^{7}$. The Blue group stops the attack.
Figure (c): amplitude $B_{1}=2.5 \cdot 10^{4}$. Disastrous defeat for
the Red group.}
\end{figure}
\subsection{Nuclear strike in course of a position conflict}
Now let a very large Red groups fights  much smaller Blue 
group. Let the Blue group has no alternative of use of non-conventional weapons 
and let it performs a nuclear strike on the Red 
group. The system of model equations is
\begin{equation}\label{nuclear1}
\frac{dB}{dt}= -rR, \hskip.5cm
\frac{dR}{dt}= -bB -H_{1}V(t,t_{1},t_{2},\mu,\nu)
\end{equation}
Results from this scenario are shown in Fig. 4. Here again
we demonstrate existence of a threshold. 
If the the nuclear strike is not massive enough (if the intensity of the
strike is below the threshold) there is no  result - the larger group wins.
 But if the strike is massive enough the situation changes and the smaller 
Blue group wins.
\begin{figure}[h]
\begin{center}
\includegraphics[angle=0,width=6cm]{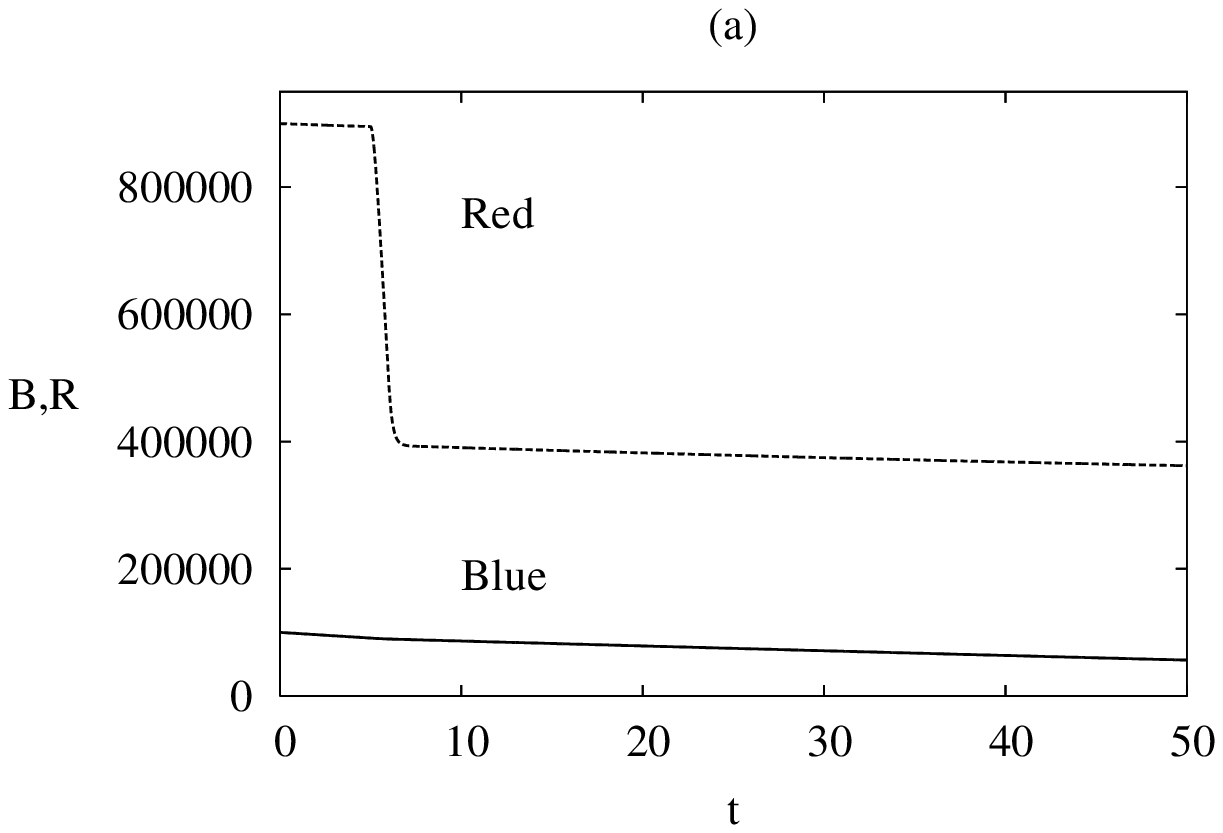}
\includegraphics[angle=0,width=6cm]{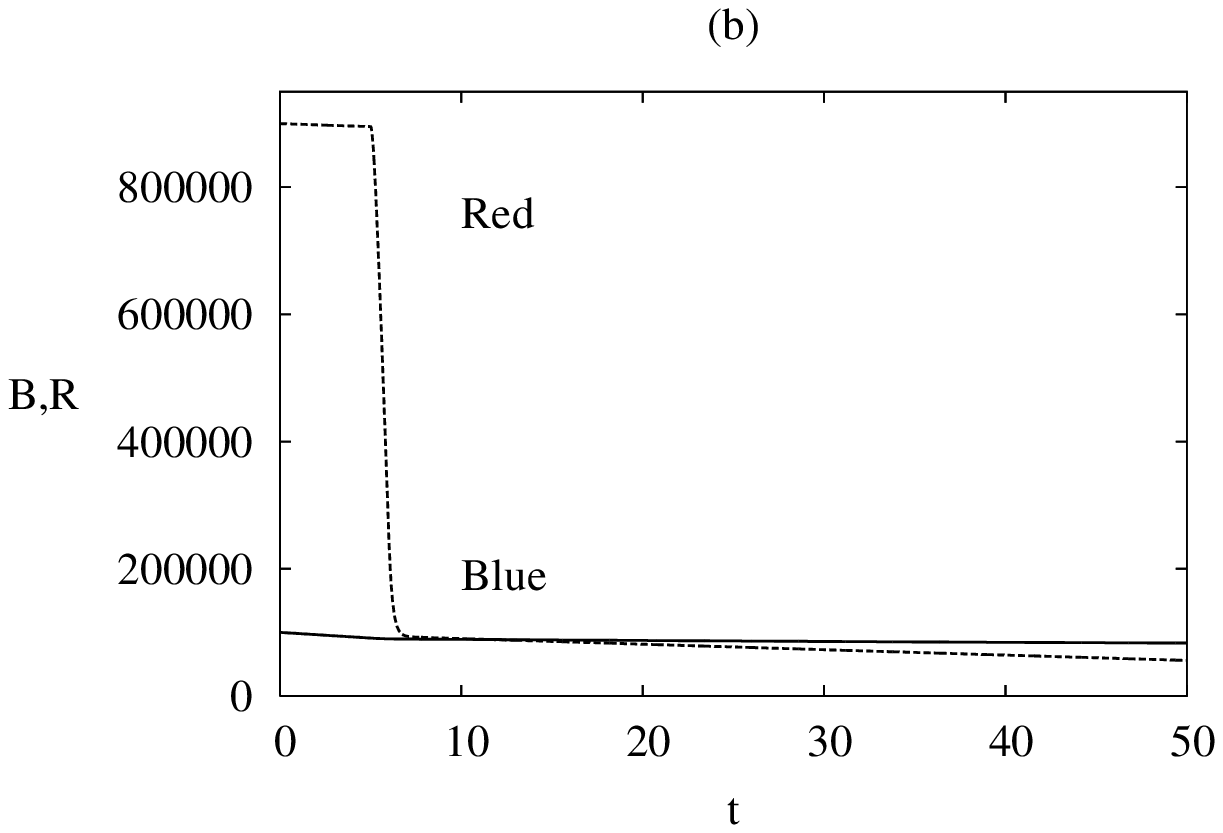}
\end{center}
\caption{Influence on nuclear strike on position conflict. Red group of 900 000 man
attacks Blue group of 100 000 man. $\mu = \nu = 5$. The strike is between
$t=5$ and $t=6$. $b=10^{-2}$, $r=2 \cdot 10^{-3}$. Figure (a): $H_{1}=
5 \cdot 10^{5}$. Inssuficient strength of the strike. The Red group
wins. Figure (b): $H_{1}=8 \cdot 10^{5}$. The strength of the
nuclear strike is sufficient. The Blue group wins.}
\end{figure}
\section{Concluding remarks}
More than 500 years ago Nicolo Machiavelli wrote that the people start a war at 
will but end it when they can. In other words it is important to know the consequences 
of the possible scenarios of a conflict. 
This can be achieved even on the basis of simple mathematical
models.  May be the most important conclusion from our results
is that the armed group which defend a teritory must not be reduced too much. 
If this happens then it is
extremely difficult and costly to fight more numerous enemy. Another conclusion is
that if the dynamics of the armed conflict become nonlinear then the accounting
for the thresholds is crucial for the success. For and example the 
inclusion of reserves must be massive or the nuclear strike must be large enough. 
Otherwise the conflict can be lost.

\begin{center}
{\sl $^{*}$ Institute of Mechanics \\ 
Bulgarian Academy of Sciences\\
Akad. G. Bonchev Str., Bl. 4 \\ 
1113 Sofia, Bulgaria}
\end{center}
\begin{center}
{\sl $^{**}$ Central Laboratory for \\
Solar Terrestial Influences of \\
Bulgarian Academy of Sciences \\
Akad. G. Bonchev Str., Bl. 3 \\
1113, Sofia, Bulgaria\\
e-mail:spanchev@phys.uni-sofia.bg} 
\end{center}
\begin{center}
{\sl $^{***}$ Faculty of Physics \\
"St. Kliment Ohridsky" University of Sofia \\
J. Bourchier 5 Blvd.\\
1164 Sofia, Bulgaria}
\end{center}
\end{document}